\documentclass[aps,prl,showpacs,twocolumn,superscriptaddress,showkeys]{revtex4}

\usepackage[T1]{fontenc}
\usepackage{times} 
\usepackage{mathptmx}
\DeclareSymbolFont{letters}{OML}{txmi}{m}{it}

\usepackage{amsmath,amssymb,graphicx,textcomp}
\usepackage{times,txfonts}

\usepackage{color}
\usepackage[normalem]{ulem}

\newcommand\diff{\mathrm{d}}

\begin{document}
\makeatletter
\def\normalsize{%
    \@setfontsize\normalsize\@xipt{12}%
    \abovedisplayskip 10\p@ \@plus2\p@ \@minus5\p@
    \belowdisplayskip \abovedisplayskip
    \abovedisplayshortskip  \abovedisplayskip
    \belowdisplayshortskip \abovedisplayskip
    \let\@listi\@listI
}%
\makeatother
\normalsize


\title{Glass  Transition in  Confined Geometry}

\author{Simon Lang}
\affiliation{Institut f\"ur Physik, Johannes Gutenberg-Universit\"at Mainz,
 Staudinger Weg 7, 55099 Mainz, Germany}

\author{Vitalie Bo\c{t}an}
\affiliation{Institut f\"ur Physik, Johannes Gutenberg-Universit\"at Mainz,
 Staudinger Weg 7, 55099 Mainz, Germany}

\author{Martin Oettel}
\affiliation{Institut f\"ur Physik, Johannes Gutenberg-Universit\"at Mainz,
 Staudinger Weg 7, 55099 Mainz, Germany}

\author{David Hajnal}
\affiliation{Institut f\"ur Physik, Johannes Gutenberg-Universit\"at Mainz,
 Staudinger Weg 7, 55099 Mainz, Germany}

\author{Thomas Franosch}
\affiliation{Institut f\"ur Theoretische Physik, Universit\"at Erlangen-N\"urnberg, Staudtstra{\ss}e~7, 91058, Erlangen, Germany}

\author {Rolf Schilling}
\affiliation{Institut f\"ur Physik, Johannes Gutenberg-Universit\"at Mainz,
 Staudinger Weg 7, 55099 Mainz, Germany}

\date{\today}

\begin{abstract}
\noindent Extending  mode-coupling theory, we elaborate a microscopic theory for the glass transition of liquids confined  between two parallel flat hard walls.
The theory contains the standard MCT equations in bulk and in two dimensions as limiting cases  and requires as input  solely  the equilibrium density profile and the structure factors of the fluid in confinement.  We evaluate the phase diagram as a function of the distance of the plates for the case of a hard sphere fluid and  obtain an oscillatory behavior
of the glass transtion line as a result of the structural changes  related to layering.

\end{abstract}





\maketitle

Dense liquids display an intriguing complex dynamical behavior upon  approaching the glass transition, as manifested for example in
the drastic slowing down of transport, the appearance of stretched structural relaxation, or the emergence of power laws at mesoscopic time
scales~\cite{Goetze:Complex_Dynamics}. For bulk systems, a scenario of the evolution of  slow complex dynamics has been presented by the mode coupling theory (MCT)
of the glass transition, many aspects of which have been confirmed  in the last two decades by experiments
and computer simulations~\cite{Goetze:Complex_Dynamics,Goetze:1999}. Yet, it is not obvious how the
approximations 
 capture the collective rearranging of the local cages which is expected to occur via increasing cooperativity.
Therefore significant
  research effort has been performed recently to
 confine the liquid to small pores, films, or tubes~\cite{Lowen:2001,Alba:2006,Workshop_Confinement:2000}, since these experiments
 may hold the key to directly unravel the  essence of the glass transition.

Computer simulations have shown that walls, 
 in general,  induce already significant changes in the static structure  and, in particular, smooth walls lead to layering~\cite{Nemeth:1999}. Second, it has been found that the dynamical features and transport properties quantified, e.g. by the diffusivity~\cite{Mittal:2008}, are strongly influenced by confinement. Here, the liquid-wall interactions play an important role: while Lennard-Jones
interactions lead to higher glass transition
temperatures by increasing the confinement~\cite{Scheidler:2000a,Scheidler:2004}, the opposite
happens for purely repulsive walls~\cite{Varnik:2002}. Interestingly, the drastic dynamical changes persist even
 in a model for an artificial pore, where the static structure remains identical to the bulk system~\cite{Scheidler:2000a,Scheidler:2004}.

Experimental results from confocal microscopy applied to colloidal suspensions
between two quasi-parallel walls reveal a decrease of the critical
packing fraction~\cite{Nugent:2007} due to confinement and a smaller mean-square displacement (MSD) parallel and close to the walls
\cite{Eral:2009}. The MSD perpendicular to the walls exhibits an oscillatory
dependence on the distance to the plates, in contrast to the MSD
parallel to the walls \cite{Nugent:2007}. A similar oscillatory behavior of the
diffusivities of a mono-disperse system of hard spheres
with packing fraction $\varphi = 0.40$ between two parallel and
flat walls has recently been found from a computer simulation,
however, for a direction perpendicular \textit{and} parallel to
the wall \cite{Mittal:2008}.
The glass transition of a single layer of  binary colloidal mixtures
\cite{Koenig:2005} displays essentially similar features as in bulk, which suggests that MCT is applicable also for the two-dimensional case~\cite{Bayer:2007,Hajnal:2009,Hajnal:2010}.

In this Letter, we extend MCT to inhomogeneous liquids confined between two parallel flat hard walls without surface roughness. The derivation employs symmetry adapted eigenfunctions and a corresponding
splitting of the current densities.
Then the theory will continuously interpolate between the glass transition in two dimensions and in bulk and relies solely on the equilibrium structure of the fluid in confinement.
As an example,  we consider a hard sphere fluid and determine the glass transition line for various distances of the plates.
In particular, we obtain the oscillatory behavior
of the glass transition line as a result of the structural changes connected to the layering.

Let us mention that MCT has already
been applied to liquids in random porous media modeled by a
quenched matrix of particles \cite{Krakoviack:2005}. This kind of MCT is quite
similar to MCT for the Lorentz model where a particle diffuses
through randomly distributed obstacles \cite{Goetze:1981} and has no
relationship to what we aim at  here. Another interesting approach is the extension of MCT to liquids in external fields~\cite{Biroli:2006}
to calculate the linear response of the intermediate scattering function to a  weak slowly varying perturbation in order to deduce a divergent dynamical length. 
Here in contrast, we consider strong confining potentials with variations even on microscopic length scales. As far as we know, MCT has never been extended to such situations, particularly to a slit, cylindrical geometries, etc. where layering occurs. 

We consider a liquid of $N$ particles confined between two
parallel flat hard walls which restrict the centers of particles to the slab between $\pm L/2$.  
The microscopic density $\rho(\vec{x},t)$
can be decomposed into suitable Fourier modes
\begin{equation}
 \rho_\mu (\vec{q},t)=\sum\limits _{n=1}^{N}\exp [i Q_\mu z_n (t)] \, \text{e}^{i \vec{q} \cdot \vec{r}_n(t)}\, ,
\end{equation}
where $\vec{r}_n(t) = (x_n(t),y_n(t))$ and $z_n(t)$ denote the position of the $n$-th
particle parallel and perpendicular to the wall, respectively.  The
discrete values $Q_\mu=2\pi \mu/L$, $\mu \in \mathbb{Z}$
account for the confined geometry and $\vec{q}=(q_x,q_y)$ refers to  the
wave vector parallel to the walls.

The walls induce a non-trivial density profile perpendicular to the confinement
 $n(z)=\left\langle \rho(\vec{x},t)\right\rangle$,
that  can be represented as a Fourier series with  Fourier coefficients
 $n_\mu=\int_{-L/2} ^{L/2}  n(z) \exp[ \text{i}Q_\mu z ]  \diff z$. 
For later purpose we also introduce the local specific volume $v(z):=1/n(z)$;
its Fourier coefficients $v_\mu$ are related to $n_\mu$ by a
convolution
 $\sum_\kappa n_{\mu-\kappa} v_{\kappa-\nu} = L^2 \delta_{\mu\nu}$.

The quantity of basic interest are the time-dependent correlation functions of density modes. Here, the  intermediate scattering function is generalized to the
infinite dimensional matrix ${\boldsymbol S}(q,t)=(S_{\mu
\nu}(q,t))$ with elements
\begin{equation}
 S_{\mu\nu}(q,t)=\frac{1}{N}\langle \delta\rho_{\mu}(\vec{q},t)^* \delta\rho_{\nu} (\vec {q},0)\rangle \, ,
\end{equation}
where $\delta \rho _\mu (\vec{q},t) = \rho _\mu(\vec{q},t)-\langle
\rho_\mu (\vec{q},t)\rangle$ denotes the  density fluctuation. Mirror reflection and time inversion symmetry (for Newtonian dynamics) implies
$n_{-\mu}=n_{\mu}$, $v_{-\mu}=v_{\mu}$, and $S_{-\mu,-\nu}(q,t)=S_{\mu\nu}(q,t)=S_{\nu\mu}(q,t)=S_{\mu\nu}(q,t)^*$, in particular
$S_{\mu\nu}(q,t)$ is real symmetric.


The continuity equation relates the densities
$\rho_\mu (\vec{q},t)$  to a corresponding  current
$\vec{j}_\mu(\vec{q},t)$. The crucial step, in contrast to bulk liquids, is now
to decompose into a component parallel and
perpendicular to the wall, $\alpha \in \{\parallel,\perp \}$,
\begin{equation}
 j_{\mu}^{\alpha}(\vec{q},t)=\sum\limits_{n=1}^{N}b^{\alpha}(\hat{\vec{q}} \cdot \dot{\vec{r}}_{n}(t),\dot{z}_n(t)) \exp[i Q_\mu z_n (t)] \, \text{e}^{i \vec{q} \cdot \vec{r}_n(t)} \, .
\end{equation}
Here, we abbreviated $\hat{\vec{q}} = \vec{q}/q$,
 and  introduced the selector $b^\alpha(x,z) = x \delta_{\alpha,\parallel} + z\delta_{\alpha,\perp}$.

Choosing $\delta \rho_\mu(\vec{q},t), \, j_\mu{^{||}}
(\vec{q},t)$ and $j_{\mu}^{\bot} (\vec{q},t)$ as distinguished variables, the
Zwanzig-Mori projection operator formalism \cite{Hansen:Theory_of_Simple_Liquids,Forster:Hydrodynamic_Fluctuations} leads to
\begin{equation}\label{eq:Phi_t}
\boldsymbol{\dot S}(q,t)+\int_{0}^{t} \diff t^\prime \boldsymbol{K}(q,t-t^\prime) \boldsymbol{S}^{-1}(q)\boldsymbol{S}(q,t^\prime)=0 ,
\end{equation}
where $\boldsymbol{S}(q) = \boldsymbol{S}(q,t=0)$ is the static correlation function. The memory kernel $\boldsymbol{K}(q,t)$ is decomposed according to
\begin{equation}\label{eq:K_t}
(\boldsymbol{K}(q,t))_{\mu\nu}=\sum_{\alpha, \beta = \perp, \parallel}b^\alpha(q,Q_\mu)\mathcal{K}_{\mu\nu}^{\alpha\beta}(q,t) b^\beta(q,Q_\nu),
\end{equation}
and its components  $\boldsymbol{\mathcal{K}}(q,t)=
(\mathcal{K}^{\alpha \beta}_{\mu \nu}(q,t))$ satisfy
\begin{equation}\label{eq:Kab_t}
\boldsymbol{\dot{\mathcal{K}}}(q,t)+\int\limits_{0}^{t}dt^{\prime} \boldsymbol{\mathcal{J}}(q)\boldsymbol{\mathcal{M}}(q,t-t^{\prime})\boldsymbol{\mathcal{K}}(q,t^{\prime})=0   .
\end{equation}
The
matrix of the static current density correlators
$\boldsymbol{\mathcal{J}}(q)=(\mathcal{J}_{\mu \nu}^{\alpha \beta}(q))$ inherits an explicit dependence on the mode index via the average density profile
\begin{equation}\label{eq:Jab_static}
 \mathcal{J}_{\mu\nu}^{\alpha\beta}(q)=\left\langle j_\mu^{\alpha}(\vec{q})^{*} j_\nu^{\beta}(\vec{q})  \right\rangle  = \frac{k_B T}{m} \frac{n_{\mu-\nu}}{n_0} \delta_{\alpha\beta}  ,
\end{equation}
where $m$ denotes the mass of the particles.

Following MCT for bulk liquids \cite{Goetze:Complex_Dynamics} the memory kernel matrix $\boldsymbol{\mathcal{M}}(q,t)= (\mathcal{M}_{\mu \nu}^{\alpha \beta} (q,t))$
can be approximated as a \emph{functional} of the density correlators
\begin{align}\label{eq:MCT}
\lefteqn{\mathcal{M}_{\mu \nu}^{\alpha \beta} (q,t)\approx\mathcal{F}_{\mu\nu}^{\alpha\beta}[\boldsymbol{S}(k,t);q] =} \\
&=
\sum_{\vec{q}_{1},\vec{q}_{2} = \vec{q}-\vec{q}_{1} }
\mathcal{V}_{\mu\mu_1\mu_2;\nu\nu_1\nu_2}^{\alpha\beta}(\vec{q},\vec{q}_{1},\vec{q_2})
S_{\mu_1\nu_1}(q_1,t) S_{\mu_2\nu_2}(q_2,t), \nonumber
\end{align}
with the vertices
\begin{align}\label{eq:Vertices}
\lefteqn{ \mathcal{V}_{\mu\mu_1\mu_2;\nu\nu_1\nu_2}^{\alpha\beta}(\vec{q},\vec{q}_{1},\vec{q}_{2})
=\frac{1}{2N}\frac{n_0^4}{L^8}\nonumber \times } \\
&\times\sum\limits_{\kappa,\kappa'} v_{\mu-\kappa}[b^\alpha(\hat{\vec{q}}\cdot
\vec{q}_{1},Q_{\kappa-\mu_2}) c_{\mu_1,\kappa-\mu_2}(q_1)+(1\leftrightarrow2)] \times\nonumber\\
&\times
[b^\beta(\hat{\vec{q}}\cdot
\vec{q}_{1},Q_{\kappa'-\nu_2})c_{\nu_1,\kappa'-\nu_2}(q_1)+(1\leftrightarrow2)]v_{\kappa'-\nu} .
\end{align}
Here, the direct correlation
functions ${\mathbf{c}}(q)=(c_{\mu \nu}(q))$ are  related to the static correlators $S_{\mu \nu}(q)$
by the inhomogeneous Ornstein-Zernike  equation~\cite{Henderson:Fundamentals_of_inhomogen
eous_fluids}
\begin{equation}\label{eq:Ornstein}
\boldsymbol{S}^{-1}(q)=\frac{n_0}{L^2}[\mathbf{v}-\mathbf{c}(q)],
\end{equation}
and $({\mathbf{v}})_{\mu \nu}= v_{\mu-\nu}$.

The MCT approximation for the memory kernel leads to a closed set
of Eqs.~\eqref{eq:Phi_t} -- \eqref{eq:Vertices} for the density correlators $S_{\mu
\nu}(q,t)$. Note that this set is valid  for any one-component liquid between two parallel
flat hard walls with \emph{arbitrary} particle-particle and
\emph{arbitrary} particle-wall interactions. These interactions
only enter into the equations via  the static quantities $n_\mu,
v_\nu$ and $S_{\mu \nu}(q)$. Let us emphasize that in contrast to bulk systems also the Fourier
coefficients $v_\mu$ of the inverse density profile enter the vertices.
   One can prove \cite{Lang:MCT_Confined} that this
set reduces to the well-known MCT equations for a bulk liquid
\cite{Goetze:Complex_Dynamics} for $L\rightarrow \infty$ and to those derived for a
two-dimensional liquid, $L \rightarrow 0$,~\cite{Bayer:2007}.
Consequently
the MCT equations~\eqref{eq:Phi_t} -- \eqref{eq:Vertices} interpolate between a two- and a
three-dimensional liquid, as required. Since the glassy dynamics is identical for both Newtonian and Brownian motion~\cite{Goetze:Complex_Dynamics}, 
our results apply also to colloidal suspensions. 

To locate the glass transition point one has to solve the
self-consistent set of equations for the non-ergodicity parameters
$F_{\mu \nu}(q):= \lim_{t \rightarrow \infty} S_{\mu
\nu}(q,t)$:
\begin{align}\label{eq:Nonergodicity}
\boldsymbol{F}(q)=&[\boldsymbol{S}^{-1}(q)+\boldsymbol{S}^{-1}(q)\boldsymbol{K}(q)\boldsymbol{S}^{-1}(q)]^{-1}\, , \\
(\boldsymbol{K}(q))_{\mu\nu}=& \sum_{\alpha,\beta =\parallel,\perp} b^\alpha(q,Q_\mu)
(\boldsymbol{\mathcal{F}}^{-1}[\boldsymbol{F}(k);q])_{\mu\nu}^{\alpha\beta}
b^\beta(q,Q_\nu). \nonumber
\end{align}


In the following we determine the glass transition line for hard spheres of diameter $\sigma$ confined between two hard walls of distance $H=L+\sigma$. 
For
the numerical solution of Eq.~\eqref{eq:Nonergodicity} one has to
truncate the Fourier indices $|\mu|\leq M$ and $q$ has to be
discretized~\cite{Lang:MCT_Confined}.
Due to the large increase in complexity
the maximum possible value has been $M=3$. In the
two-dimensional limit $H/\sigma \rightarrow 1$ only the lowest mode contribute, $S_{\mu
\nu}(q,t) \rightarrow S_{00}(q,t) \delta_{\mu \nu}
\delta_{\nu 0}$. Already for $H/\sigma \gtrsim 2$
higher order modes $|\mu|>0$ have to be taken into account,
particularly for $H/\sigma \gg 1$. Therefore we have also performed a subsequent
diagonalization approximation (DA), where only the diagonal
elements $S_{\mu \mu} (q), \; c_{\mu \mu}(q), \; F_{\mu
\mu}(q)$ and $\mathcal{F}_{\mu \mu}^{\alpha \alpha} [\{F_{\mu
\mu}(k)\};q]$ have been taken as nonzero, allowing to handle $M
= 10$.


\begin{figure}
\includegraphics[width=\columnwidth]{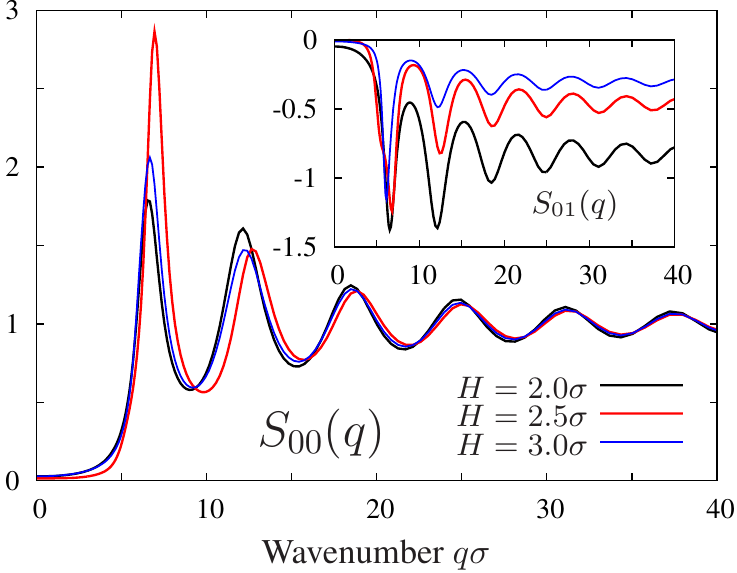}
\caption{\label{fig:Static_correlators} Static correlators $S_{00}(q)$ for $\varphi = 0.42$ for different distances $H$
of the plates. The first sharp diffraction
peak varies nonmonotonically; lowest for $H=2.0\sigma$, highest for $H=2.5\sigma$ and intermediate for $H=3.0\sigma$.
Inset: The corresponding off-diagonal structure factor $S_{01}(q)$ is negative and becomes smaller upon increasing $H$.}
\end{figure}

The static input quantities were obtained by first calculating  the density
profile $n(z)$ within fundamental measure theory~\cite{Hansen-Goos:2006}. These results are used  as input into the
inhomogeneous Ornstein-Zernike equation which is then solved with the
Percus-Yevick closure relation~\cite{Botan:2009}.
The depletion forces give rise to a strongly enhanced density profile in the vicinity of walls, 
which results in the characteristic layering.
The principle peak in the static structure factor $S_{00}(q)$ at $q\sigma \approx 2 \pi$ is
higher for the half integer distance
$H/\sigma = 2.5$ than for both integer distances $H/\sigma =2.0$ and
3.0, see Fig.~\ref{fig:Static_correlators}. Consequently the static correlations appear reduced in the case that
a few  monolayers just fit between the walls, and one anticipates this layering effect to have drastic influences on the glass transition.

\begin{figure}
\includegraphics[width=\columnwidth]{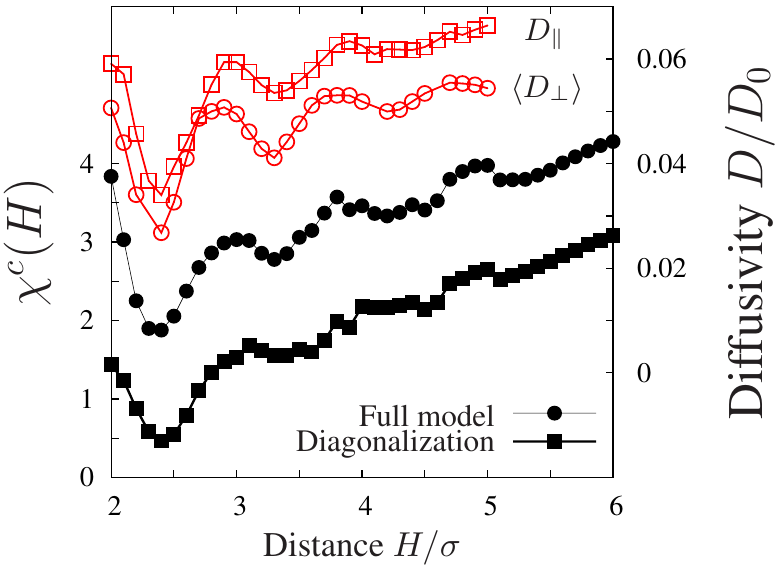}
\caption{Critical parameter
$\chi_c(H;\varphi = 0.42)$ for $M = 3$, with DA  and
without DA, and  the dimensionless
diffusivities $D_{||}$ and $D_\bot$, respectively, taken from
\cite{Mittal:2008} for $\varphi = 0.40$.}
\label{fig:chi_critical}
\end{figure}

The maximum packing fraction $\varphi := \pi n\sigma^3/6$ ($n = n_0/H$), for which we were able to reliably  calculate the static
input quantities  has been for  $\varphi = 0.42$, so far,  not large enough to induce a transition. Therefore we
mimic the increase of the static correlations by an additional
multiplicative control parameter $\chi$  in front of the MCT functional
$\mathcal{F}_{\mu \nu}^{\alpha \beta}$. Let $\varphi _c(H)$ be the
critical packing fraction for $\chi =1$. Then, if $\varphi $ is below
and close to $\varphi _c(H)$ the vertices vary linearly with
$\varphi$, and  we therefore expect that the $H$-dependence of
$\varphi_c(H)$ is qualitatively well described by the $H$-dependence
of the critical parameter $\chi_c(H;\varphi)$. Figure~\ref{fig:chi_critical} depicts
$\chi_c(H;0.42)$ obtained with and without DA for $M = 3$; both
graphs strongly resemble, particularly in the region $2.0 \leq H/\sigma
\leq 4.0$ where $\chi_c$ has larger variation in contrast to $H/\sigma >4.0$.
The larger absolute values of $\chi_c$ without DA may originate
from  a partial cancelation in the vertices.
 Increasing the cut-off $M$ in the DA, i.e. taken into account higher order modes, we found that
$\chi_c(H;0.42)$
does
not change  qualitatively  (not shown).

As argued above, we expect for $\varphi_c(H)$ qualitatively a
similar $H$-dependence as we have found for $\chi_c(H;0.42)$. Now
assume  $\varphi$ fixed below $\varphi_c(H)$. Then, the distance
$(\varphi_c(H)-\varphi)$ to the glass transition point is maximum
(minimum) for those $H$ for which the $\chi_c(H;\varphi)$ is maximal (minimal).
MCT  predicts for the diffusivity $D^{\text{MCT}}(H,\varphi) \sim [\varphi_c(H)-\varphi]^{\gamma(H)}$ where $\gamma(H)$ does not depend sensitively on the
type of the liquid and is close to two. Therefore assuming only a weak dependence of $\gamma$ on $H$, MCT predicts a  maximum
diffusivity at $(H/\sigma)_{\text{max}}^{\text{MCT}}\approx 2.0, \;
\approx 3.1, \; \approx 3.8$ and $\approx 5$. These values are
surprisingly close to the corresponding values from the simulation,  Fig.~\ref{fig:chi_critical},
particularly for $\langle D_\bot \rangle^{\text{sim}}$. Note that
the
variation of the diffusivity with $H$ is much less pronounced for
$H/\sigma> 4.0$ than for $H/\sigma <4.0$ which is reproduced by MCT, as well.

In conclusion, we have generalized MCT to liquids confined between two
parallel flat walls for arbitrary particle-particle and
particle-wall interactions. The theory differs in several important aspects from corresponding one in bulk and in the plane. First,
the walls exchange momentum with the fluid implying that the intermediate scattering function is no longer diagonal in the wavenumber.
Second, the theory  requires a splitting of the currents in order to capture the limits $H\to \sigma$ and $H\to\infty$. Furthermore  the theory
requires explicitly the  density profile induced by the confinement, in addition to the generalized structure factors, which renders the glass transition line sensitive
to layering effects. We have exemplified the predictive power of the theory by reproducing the non-monotonic behavior of the diffusivity for hard spheres on the
confinement. Particularly, its different behavior for integer and half-integer values of  $H/\sigma$ can be interpreted as follows. If $H/\sigma$
 equals an integer $r$ then exactly $r$
monolayers fit into the slit. If the monolayers were perfectly
flat, they could slide. In order to get a structural arrest
$\varphi$ has to be increased. If $H/\sigma$ deviates stronger from $r$
the monolayers become rougher and penetrate into each other which
reduces the particle's mobility and hence the packing fraction
for a structural arrest.  This interpretation is consistent with the $H$ dependence of the principal peak of $S_{00}(q)$, see Fig.~\ref{fig:Static_correlators}.

The $H$-dependence of $D$ has also been related to that of static quantities such as the excess entropy $s^{\text{ex}}$ and the  available volume fraction $\overline{p_0}$. Knowledge of the bulk diffusivity $D_{\text{bulk}}$ then yields $D(H,\ldots) \approx D_\text{bulk}(x(H),\ldots)$ for $x = s^{\text{ex}}, \overline{p_0}$ provided $D$ is not too small~\cite{Goel:2009}. In MCT $x$ corresponds to $\varphi_c(H)$ and, if $\gamma(H)$ is almost constant, a similar relation $D^{\text{MCT}}(H,\ldots) \approx D^\text{MCT}_\text{bulk}(\varphi_c(H),\ldots)$ holds. Calculating $s^\text{ex}(H)$ and $\overline{p_0}(H)$ from fundamental measure theory, we 
find that $s^\text{ex}$ and $\overline{p_0}$ are not constant on the glass transition line. However, the $H$ dependence of $\varphi_c$ is in phase with that of $s^\text{ex}$ and $\overline{p_0}$ demonstrating the importance of the latter for the glass transition.  

Recently,
it has been shown that bulk binary mixtures of hard spheres~\cite{Foffi:2003,Goetze:2003} or hard disks~\cite{Hajnal:2009} display an intriguing
dependence on composition. In confinement these mixing effects compete with the layering induced by the walls and should give rise to  an even richer
phase behavior. 
It is instructive to also compare the glass transition line with the random close packing values. 
For hard disk mixtures in the bulk  both lines exhibit a striking similarity~\cite{Hajnal:2009}, and  the random close packing of a binary mixture of hard spheres in confinement displays an oscillatory dependence on the distance of the walls~\cite{Desmond:2009}.
Thus, it would be interesting to extend  MCT  also to mixtures in confinement. In addition, 
the theory presented here constitutes a promising framework to rationalize the numerous glass-forming systems in simple confinements that have been investigated in the recent past~\cite{Workshop_Confinement:2000}.
Furthermore, it is a challenge for the future to extend MCT also to pore models for which the static properties  are uneffected  whereas the dynamical features sensitively depend on the confinement~\cite{{Scheidler:2000a,Scheidler:2004}}. 

\acknowledgments
We thank K. Binder and B. Schmid for helpful discussions and  J. Mittal and T. Truskett for providing their data on the diffusivity.
We  acknowledge support via the DFG research unit FOR-1394 (T.F.) and SFB-TR6 (N01, M.O. and V.B.).


\end{document}